\documentclass{Interspeech2024}
\usepackage{algorithmic}
\usepackage{microtype}
\usepackage{graphicx}
\usepackage{booktabs} % for professional tables
\usepackage{hyperref}
\usepackage{amssymb}
\usepackage{mathtools}
\usepackage{amsthm}
\usepackage{wasysym}
\usepackage{graphicx}
\usepackage[ruled,vlined]{algorithm2e}
\usepackage{caption}
\usepackage{subcaption}
\usepackage{xcolor}
\interspeechcameraready

\title{Optimal Transport Maps are Good Voice Converters}

\name[affiliation={* 1,2}]{Arip}{Asadulaev}
\name[affiliation={* 3}]{Rostislav}{Korst}
\name[affiliation={2}]{Vitalii}{Shutov}
\name[affiliation={4,1}]{Alexander}{Korotin}
\name[affiliation={3}]{Yaroslav}{Grebnyak}
\name[affiliation={5}]{Vahe}{Egiazarian}
\name[affiliation={4,1}]{Evgeny}{Burnaev}

\address{
  $^1$AIRI,
  $^2$ITMO
  $^3$MIPT
  $^4$Skoltech
  $^5$Yandex
  }
\email{aripasadulaev@airi.net}


\DeclareMathOperator*{\argsup}{arg\,sup}

\begin{document}

\maketitle

\begin{abstract}
Recently, neural network-based methods for computing optimal transport maps have been effectively applied to style transfer problems. However, the application of these methods to voice conversion is underexplored. In our paper, we fill this gap by investigating optimal transport as a framework for voice conversion. We present a variety of optimal transport algorithms designed for different data representations, such as mel-spectrograms and latent representation of self-supervised speech models. For the mel-spectogram data representation, we achieve strong results in terms of Fréchet Audio Distance (FAD). This performance is consistent with our theoretical analysis, which suggests that our method provides an upper bound on the FAD between the target and generated distributions. Within the latent space of the WavLM encoder, we achived state-of-the-art results and outperformed existing methods even with limited reference speaker data.
\end{abstract}
% Using various automatic metrics, we evaluated the quality of the voice conversion models. Our method outperforms the previously proposed voice conversion approaches and exhibits high quality results while being light.The clarity of the proposed learning process enables for a theoretical analysis.
\vspace{-10pt}
\section{Introduction}
The goal of VC is to generate a modified voice that maintains the linguistic content of the source speaker while adopting the prosody and vocal characteristics of a target speaker~\cite{walczyna2023overview}. It has various applications including voice modification~\cite{eskimez2022personalized}, singing voice conversion~\cite{ huang2023singing}, and privacy~\cite{Qian2019}. The most common case is non-parallel VC, i.e. when the speech content is different for the source and target speaker datasets. Existing VC methods such as StarGAN-VC~\cite{goodfellow2014generative, Li2021}, Diff-VC~\cite{ popov2021diffusion} and kNN-VC~\cite{baas2023voice} deliver impressive results. But these models have an drawbacks, such as complex training procedures, significant computational resources required for inference or large quantity of target speaker data (\wasyparagraph\ref{sec:background}).\\
There has been recent interest in the use of neural network-based optimal transport (OT) maps for generative modelling in high-dimensions~\cite{ korotin2019wasserstein,  korotin2023neural, asadulaev2022neural}. However, the potential of optimal transport maps in voice conversion has not been comprehensively explored. In this paper, we fill this gap by proposing variants of OT methods for voice conversion.
\begin{itemize}
    \item For mel-spectogram representation we present a neural network-based OT approach called NOT-VC (\wasyparagraph\ref{sec:mel-method}). This method proves to be resource-efficient in terms of inference and simpler in terms of training compared to its predecessors as DiffVC and StarGAN-VC respectively. We justify our approach theoretically by analyzing the recovered maps and show that the learned map upper bounds the FAD~\cite{FAD} between real and generated data (\wasyparagraph\ref{sec:cnot}). Furthermore, we extend this method to create an Extremal NOT~\cite{gazdieva2023extremal} named XNOT-VC (\wasyparagraph\ref{sec:xnot}) and demonstrate state-of-the-art (SOTA) performance according to FAD. 
    \item For voice conversion within the latent space of the WavLM~\cite{chen2022wavlm} model (\wasyparagraph\ref{sec:fmnot}), we propose an OT-based Flow-Matching~\cite{lipman2022flow, tong2023improving} approach named FMVC. This method is lightweight and effectively circumvents the limitations of the current best-performing any-to-any VC method named kNN-VC~\cite{baas2023voice}, providing SOTA results even with limited target data (\wasyparagraph\ref{sec:unit-experiments}).\\
\end{itemize}
In our paper, we explore the applicability of optimal transport to non-parallel VC. The main \underline{contribution} is the development and theoretical justification of several OT methods that achieve impressive results and avoid the limitations of existing VC approaches.

\section{Background and Related Work}
\label{sec:background}
 \subsection{Optimal Transport}
\label{sec:ot_background}
%, and in Section 3 we propose our methods for VC using OT.
Optimal transport is a mathematical tool designed to minimize the cost of moving mass between distributions. Suppose there are two probability distributions $\mu$ and $\nu$ over measurable spaces $\mathcal{X}$ and $\mathcal{Y}$ respectively, where $\mathcal{X},\mathcal{Y}\subset\mathbb{R}^{D}$. We want to find a measurable map $T: \mathcal{X}\rightarrow\mathcal{Y}$ such that mapped distribution is equal to the target $\nu$, $T_\#\mu = \nu$. For a cost function $c:\mathcal{X}\times\mathcal{Y}\rightarrow\mathbb{R}$, the \textit{OT problem} between $\mu,\nu$ is 
\begin{equation}
    \inf_{T_\sharp\mu = \nu}\int_\mathcal{X} c(x, T(x))\, d\mu(x).
    \label{eq:monge-ot}
\end{equation}
In the \textit{Kantorovich OT formulation}~\cite{kantorovitch1958translocation}, we are seek for a probability measure $\pi$ over $\mathcal{X}\times\mathcal{Y}$ where marginals over $\pi$ satisfying $\pi_{x}=\nu$ and $\pi_{y}=\mu$, respectively \cite[\wasyparagraph 1]{villani2008optimal}. Denote the set of all such $\pi$ by $\Pi(\mathcal{X}, \mathcal{Y})$, the problem is to find an optimal transportation plan $\pi^*$ that minimizes
\begin{equation}
\inf_{\pi\in\Pi(\mu,\nu)}\int_{\mathcal{X}\times\mathcal{Y}}c(x,y)d\pi(x,y).
\label{eq:ot-kantorovich}
\end{equation}\\
For the cost function $c(x,y)=\frac{1}{2}|x-y|_{2}^{2}$, the solution value of (\ref{eq:monge-ot}, \ref{eq:ot-kantorovich}) is called Wasserstein-2 ($\mathbb{W}_{2}$) distance \cite[\wasyparagraph 1]{villani2008optimal}. To solve the OT problem in practice, different discrete solvers such as Sinkhorn can be used \cite{peyre2019computational}. For continuous OT in high dimensions, \textit{maximin} form of (\ref{eq:monge-ot}) with a neural approximators is used~\cite{ korotin2019wasserstein,  korotin2023neural, asadulaev2022neural}. Recently the Flow-Matching (FM) approach \cite{lipman2022flow} with a simple training
objective that regress onto a target vector field was proposed. For the target vector field that corresponds to an OT displacement interpolant, this methods provides a straight paths between points, and approximates OT map \cite{lipman2022flow, pooladian2023multisample}. 
\subsection{Voice Conversion}
\label{sec:background}
Most VC methods operate with the mel-spectrogram representations~\cite{walczyna2023overview}. These representations convert raw audio signals into a 2D time-frequency representation that displays the frequency of speech over time. After conversion, vocoder models are typically used to transform these mel-spectrograms back into raw audio. Despite their effectiveness, this VC methods have specific drawbacks related to their distinct approaches.  Diffusion-based models as Diff-VC~\cite{  popov2021diffusion} require substantial computational resources due to the \underline{numerous steps} involved in inference. On the other hand, generative adversarial networks-based (GANs)~\cite{goodfellow2014generative} methods, such as StarGAN-VC~\cite{goodfellow2014generative,  Li2021}, \underline{optimize multiple losses} which – when combined with different coefficients – complicate the optimization and hyperparameter tuning process. A more detailed discussion on related works in provided in Appendix (\wasyparagraph\ref{sec:baselines_appendix})\\
Within the latent space of the speech representation model, unit-selection methods have been introduced. These methods convert speech between the pair of the source and target speakers by replacing each frame of the source speech with its selected correspondence in the target representation. A strong result was achieved by the kNN-VC~\cite{baas2023voice} method. It was shown that replacing the source speech frames with their nearest neighbors in the target speech provides a SOTA result for any-to-any conversion. Despite its advantages, this method requires \underline{extensive recordings} (5-10 minutes) of the target speaker for accurate speech conversion, see Figure 2 in \cite{baas2023voice}. This is an obvious limitation, because if the recording of the target speech is too short, we simply won't be able to find where to match the phones and biphones present in the source utterance \cite[\wasyparagraph 5.2]{baas2023voice}.
\section{Voice Conversion with Optimal Transport}
\label{sec:mel-method}
\subsection{Conditional Neural Optimal Transport}
\label{sec:cnot}
\begin{figure}[h!]
    \centering
    \includegraphics[width=7cm]{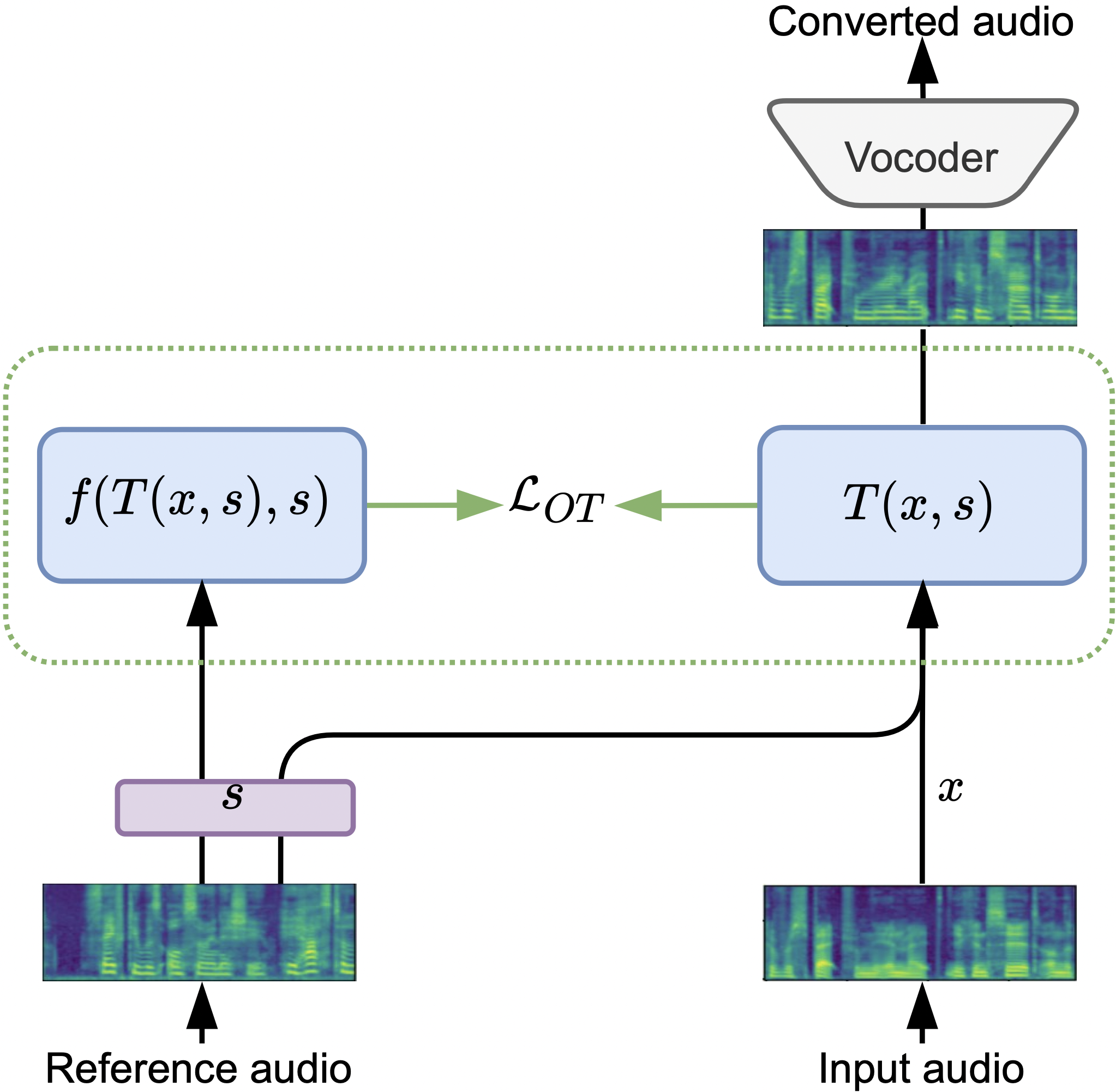}
    \caption{First, the mel-spectrogram of the source speaker is fed into the map $T$, while the reference is fed into the speaker encoder and also used as input for $T$ and $f$ (during training). The map $T$ outputs the converted spectrogram, which is transformed back into raw audio by the vocoder..}
    \label{fig:mel_model}
\end{figure}

%The typical setting for speech conversion, is voice style transfer across multiple domains, wherein each domain is a unique source and target speaker. Given the nature of this task, conditioned generator networks are utilized in VC methods~\cite{ Li2021}. 
To solve OT problems in high dimensions, the most popular approaches~\cite{ korotin2019wasserstein,  korotin2023neural, asadulaev2022neural} consider \textit{maximin} form of (\ref{eq:monge-ot}):
\begin{equation}
 \max_{f}\min_{T}\int_{\mathcal{X}}c\big(x,T(x)\big)-f\big(T(x)\big)d\mu(x)+\int_{\mathcal{Y}}f(y)d\nu(y).
\label{eq:dual-ot}
\end{equation}
Where potential $f$ is \textit{Lagrangian multiplier}~\cite{OTNotes}, that aims to ensure that the generated distribution by $T$ matches the target distribution $\nu$, and penalizes the former if it does not. Within neural approximations for $T$ and $f$, this type of methods are called Neural Optimal Transport (NOT)~\cite{  korotin2023neural}. To build OT map for VC across multiple domains, we propose a conditional reformulation of the NOT method~\cite{korotin2021neural}. Let denote distributions, of source speaker $\mu$, target speaker$\nu$, and reference speaker $\eta$ over measurable spaces $\mathcal{X}$, $\mathcal{Y}$ and $\mathcal{S}$ respectively, where $\mathcal{X},\mathcal{Y}, \mathcal{S}\subset\mathbb{R}^{D}$. We want to find the transport map $T$ that for any $s$ map conditioned distribution  $\mu(\cdot|s)$ to the respected one $\nu(\cdot|s)$. Formally we can write \underline{our objective} as:
\begin{align}
\mathcal{L}_{OT}=\displaystyle\max_{f}\min_{T}\displaystyle\int_{\mathcal{X}\times\mathcal{S}} (c\big(x,T(x,s)\big)- \nonumber\\
f\big(T(x,s), s\big))d\mu(x,s)+\displaystyle\int_{\mathcal{Y}\times\mathcal{S}} f(y,s)d\nu(y,s).\label{eq:cond_dual-ot}
\end{align}
Now, the \textit{Lagrangian multiplier} $f$, aims to ensure that the generated distribution $T_\sharp \mu(\cdot|s)$ matches the target distribution $\nu(\cdot| s)$ for any given $s$. The solution to the problem (\ref{eq:cond_dual-ot}) can be practically carried out by using neural networks $T_{\theta}:\mathbb{R}^{D}\times\mathbb{R}^{S}\rightarrow\mathbb{R}^{D}$ and $f_{\psi}:\mathbb{R}^{D}\rightarrow\mathbb{R}$ to parameterize $T$ and $f$ respectively. In practice, the speaker encoder model $E_{\omega}$ to embed the reference speaker $s$ is used. The entire \underline{optimization procedure} can be found in Appendix (\wasyparagraph~\ref{sec:algorithm_appendix}), the visual illustration is given in Figure~\ref{fig:mel_model}.\\ 
Given a pair ($\hat{f},\hat{T}$) that approximately solves equation (\ref{eq:cond_dual-ot}), it is natural to question the quality of the recovered map $T$. To answer this, we provide a bound on the difference between the optimal $T^*$ that maps into the target and $T$. \\
\textbf{Theorem 1.} \textit{(informal). Assume that there exists a unique deterministic OT plan for quadratic cost between $\mu$ and $\nu$, i.e., $\pi^*$ for $T^{*}:\mathbb{R}^{D}\times\mathbb{R}^{S}\rightarrow\mathbb{R}^{D}$. Assume that $\hat{f}$ is $\beta$-strongly convex $(\beta>0)$ and $\hat{T}:\mathbb{R}^{D}\times\mathbb{R}^{S}\rightarrow\mathbb{R}^{D}$. Then the map obtained by minimizing (\ref{eq:cond_dual-ot} upper bounds the FAD between the $\nu$ and distribution generated by $T$}. More details and proof are given in Appendix (\wasyparagraph\ref{sec:analysis}).
\vspace{-3mm}
\vspace{-6mm}
% \begin{algorithm}[t!]
% \SetInd{0.5em}{0.3em}
%     {\SetAlgorithmName{Algorithm}{empty}{Empty}
%         \SetKwInOut{Input}{Input}
%         \SetKwInOut{Output}{Output}
%         \Input{conditional distributions $\mu(\cdot|s),\nu(\cdot|s)$ for all $s$ accessible by samples, map $T_{\theta}$, potential $f_{\psi}$, cost $c(x,y)$; number of inner iterations $K_{T}$.}
%         \Output{approximate OT map  $(T_{\theta})_{\#}\mu(\cdot|s)=\nu(\cdot|s)$}}
%         \Repeat{not converged}{
%             sample $s\sim\eta$; $x\sim\mu(\cdot|s)$, $y\sim \nu(\cdot|s)$;
%             $\mathcal{L}_{f}\leftarrow  \frac{1}{|Y|}\sum\limits_{y\in Y}f_{\psi}(y,s) - \frac{1}{|X|}\sum\limits_{x\in X}f_{\psi}\big(T_{\theta}(x,s),s\big)$;
%             update $\psi$ by using $\frac{\partial \mathcal{L}_{f}}{\partial \psi}$ to maximize $\mathcal{L}_{f}$;
            
%             \For{$k_{T} = 1,2, \dots, K_{T}$}{
%                 sample $s\sim\eta$; $x\sim\mu(\cdot|s)$, $y\sim \nu(\cdot|s)$; ${\mathcal{L}_{T}\leftarrow\frac{1}{|X|}\sum\limits_{x\in X}\big[c
%                 \big(x, T_{\theta}(x,s)\big)- f_{\psi}\big(T_{\theta}(x,s),s\big)\big]}$;
%             update $\theta$ by using $\frac{\partial \mathcal{L}_{T}}{\partial \theta}$ to minimize $\mathcal{L}_{T}$;
%             }
%         }
%         \caption{Voice Conversion with Neural OT}
%         \label{algorithm:main}
% \end{algorithm}\\
\subsection{Extremal Conditional Optimal Transport}\vspace{-1mm}
\label{sec:xnot}
Usually, samples in the target domain may be noisy or too different from the source, especially in non-parallel VC setups. Thus, in some cases, it would be beneficial to not use only the part of the target distribution.
For this, the extremal transport maps~\cite{gazdieva2023extremal} can be used. Extremal transport performs outlier detection in the target space by ignoring samples that are too different from the source distribution with respect to the given cost. In this formulation, we learn the \textit{$T$ that maps only into the part} of the target distribution that reduces the cost. To obtain the extremal formulation from NOT-VC we simply need to add $f\leq 0$ constraints over $f$ and multiply $f(y,s)$ by a weight parameter $w\geq 1$. This results in the following optimization:
\begin{eqnarray}
& \mathcal{L}_{EOT}=\displaystyle\max_{{\color{purple}f\leq 0}}\min_{T}\displaystyle\int_{\mathcal{X}\times\mathcal{S}} (c\big(x,T(x,s)\big)- \\\nonumber
& f\big(T(x,s), s\big))d\mu(x,s)+{\color{purple}w}\displaystyle\int_{\mathcal{Y}\times\mathcal{S}} f(y,s)d\nu(y,s).
\label{eq:extremal_cond_dual-ot}
\end{eqnarray}
This method can be seen as a tool for finding the \textit{nearest neighbors} of the input samples to the target, according to the cost function, (see Figure 1 in \cite{gazdieva2023extremal}). Importantly, the parameter $w$ in this formulation controls the closeness of the generated samples to the input ones, see ~\cite{gazdieva2023extremal} for more details. %To obtain an \textit{extremal version} of the Algorithm~\ref{algorithm:main}, the only changes highlighted in {\color{purple} purple} in (\ref{eq:extremal_cond_dual-ot}) should be applied.  %Similar to (\ref{eq:cond_dual-ot}), this formulation can be efficiently computed using neural network approximators for $T$ and $f$. 

% Full identity consistency: a further constraint on the input preservation, the identity-mapping loss is used:
% \begin{equation}
%     \mathcal{L}_{i d}=\mathbb{E}_{(\boldsymbol{x}, c) \sim P(\boldsymbol{x}, c)}[|G(\boldsymbol{x}, c)-\boldsymbol{x}|]
% \end{equation}

\subsection{Flow-Matching Optimal Transport}
\label{sec:fmnot}
\begin{figure}[h!]
    \centering
    \includegraphics[width=7cm]{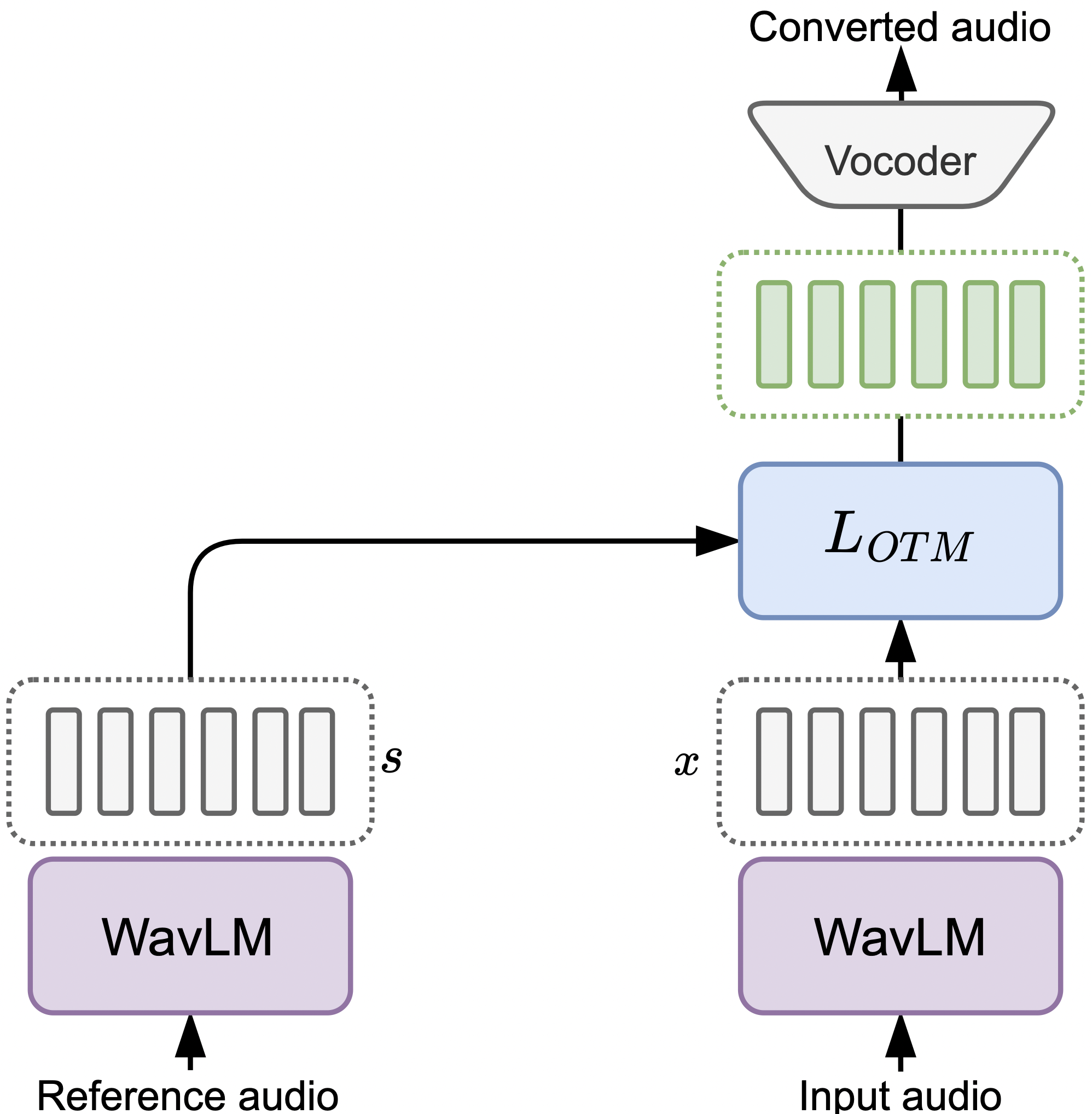}
    \caption{Wav audio is fed into the WavLM. At the same time, the reference is fed into the WavLM. Then the OT matching $\mathcal{L}_{OTM}$ (SinkVC or FMVC) converts the voice into the given latent representations. After inference, the results are transformed back into raw audio using the vocoder.}
    \label{fig:Wav_model}
\end{figure}
It has been shown that self-supervised models extract representations where nearby features have similar phonetic content. We propose a method that uses a self-supervised speech representation model to encode the audio  \cite{chen2022wavlm}. This approach provides an \textit{encoder-converter-vocoder} structure. Firstly, we extract self-supervised representations of the source and reference speech. Then, conversion to the target speaker takes place, by replacing each frame of the source representation by frame from the reference via OT map. Finally, a pre-trained vocoder synthesizes audio from the converted representation. 
We solve a matching OT problem (\ref{eq:ot-kantorovich}) between source and target bag-of-vectors, using the Sinkhorn algorithm \cite{cuturi2013sinkhorn}. The OT plan $\pi$ provides optimal pairs between source and reference speaker representations. This plan already provides a solution to the VC problem. By replacing each source frame with its corresponding frame given by the plan, we can solve the VC problem in real-time. We call this approach SinkVC.\\
However, SinkVC, as well as kNN-VC, provides a \underline{discrete} solution which may prove inefficient in cases where the amount of reference speech is limited. Denoting pairs given by a plan as $(x_0, x_1)\sim \pi$, we can train an FM algorithm (\ref{eq:FM}) that generates a \underline{continuous} OT map for VC (FMVC). OT-based Flow Matching predicts a target velocity field, which is defined by the OT interpolation between points. Formally, using an OT plan given by Sinkhorn, we sample pairs $(x_0, x_1)\sim \pi$ \cite{pooladian2023multisample, tong2023improving} while minimize:
\begin{equation}
L_{OTM}(\theta)=\mathbb{E}_{t,(x_0,x_1)\sim \pi}||v_\theta(t,x_t)-(x_0-x_1)||^{2}. 
\label{eq:FM}
\end{equation}
FMVC produces a flow induced by a neural velocity field $(v_t,\theta), t \in [0,1]$, constructing probability paths between individual data samples. The solution generates individual straight paths between source and target pairs. It has been shown that the obtained flow asymptotically approximates an OT map see Theorem 4.2 in \cite{pooladian2023multisample}. A visual illustration is shown in Figure \ref{fig:mel_model}.

\section{Experiments with Mel-Spectogram Audio Representation}
\label{sec:experiments}
\begin{table}[h!]
\centering
%\small
\begin{center}
%\begin{small}
\begin{tabular}{l|lllll}
\hline
Model      & FAD $\downarrow$    & EER $\uparrow$   & pMOS $\uparrow$ \\ \hline
AutoVC          & 10.9  & 0.18   & 3.51  \\
StarGANv2-VC    & 1.18  & \underline{0.24} & \textbf{4.29}  \\
DiffVC         & 1.34  & \textbf{0.38}  & 3.83 \\\hline
NOT-VC (Ours)   & \underline{1.04}  & 0.18    & \underline{4.18}  \\
XNOT-VC (Ours)  & \textbf{0.87}  & 0.17   & 3.85  \\
\hline
\end{tabular}
%\end{small}
\end{center}
\caption{Many-to-many conversion results on the VCTK dataset. The symbol $\uparrow$ indicates that a higher score is better, while $\downarrow$ indicates that a lower score is better.}
\label{tab:table-1}
\end{table}
\begin{table}[h!]
\centering
\begin{center}
%\small
%\begin{small}
\begin{tabular}{l|lllll}
\hline
Model      & FAD $\downarrow$    & EER $\uparrow$   & pMOS $\uparrow$ \\ \hline
StarGANv2-VC     & \underline{1.34}  & 0.076          & \textbf{4.22}  \\
DiffVC           & 1.46           & \textbf{0.380}  & 3.56 \\\hline
NOT-VC (Ours)     & 1.46           & \underline{0.078}   & \underline{4.00}  \\
XNOT-VC (Ours)    & \textbf{1.33}  & 0.043          & 3.74 \\
\hline
\end{tabular}
%\end{small}
\end{center}
\vskip -0.1in
\caption{Any-to-many conversion results on the VCTK dataset. The symbol $\uparrow$ indicates that a higher score is better, while $\downarrow$ indicates that a lower score is better.}
\label{tab:table-2}
\end{table}
%The goal of such a setup is to test the ability of voice conversion models to convert the voice of the previously unseen speaker to that of the speakers on which the model was trained. 
% \begin{figure*}[h!]
%      \centering
%      \begin{subfigure}[b]{0.3\textwidth}
%          \centering
%          \includegraphics[width=\textwidth]{AISTATS2023PaperPack/Images/Input.png}
%          \caption{Input sample $x$}
%          \label{fig:y equals x}
%      \end{subfigure}
%      \hfill
%      \begin{subfigure}[b]{0.3\textwidth}
%          \centering
%          \includegraphics[width=\textwidth]{AISTATS2023PaperPack/Images/Reference.png}
%          \caption{Reference sample $s$}
%          \label{fig:three sin x}
%      \end{subfigure}
%      \hfill
%      \begin{subfigure}[b]{0.3\textwidth}
%          \centering
%          \includegraphics[width=\textwidth]{AISTATS2023PaperPack/Images/Results.png}
%          \caption{Results of the transport map}
%          \label{fig:five over x}
%      \end{subfigure}
%         \caption{Three simple graphs}
%         \label{fig:three graphs}
% \end{figure*}
%\subsection{Datasets}
%\subsection{Mel-Spectogram Representation}
\subsection{Dataset and Baselines}
\label{sec:baselines}
In this section we provide experiments using the common multi-speaker VCTK dataset ~\cite{veaux2016superseded}. This dataset contains recorded speech data from 109 native English speakers, amounting to approximately 44 hours of speech. Similar to StarGANv2-VC, we used 20 randomly chosen speakers from the dataset for training. Data prepossessing details are given in Appendix (\wasyparagraph\ref{sec:details_data}).\\
We compared our method with the range of methods such as \textit{AutoVC}~\cite{Qian2019}, a 127M parameters pre-trained \textit{Diff-VCTK}~\cite{popov2021diffusion}, and  \textit{StarGANv2-VC}~\cite{Li2021}. In all experiments, we used pre-trained modes provided by the authors. The detailed explanation of experiments is given in Appendix (\wasyparagraph\ref{sec:baselines_appendix}). 
\subsection{Settings} 
For the transport map $T_\theta$, a UNET network with 46 M parameters was utilized. Meanwhile, for the potential $f_\psi$, a ResNet model with 55.1 M parameters was employed. AdamW~\cite{AdamW} was selected as the optimizer for both the map $T$ and the potential $f$. Our methods were trained using Algorithm~\ref{algorithm:main} given in the Appendix, incorporating a quadratic cost $c(x,y)=\frac{1}{2}|x-y|_{2}^{2}$. The complete details of our training are available in Appendix~(\wasyparagraph\ref{sec:details_models}). For a fair comparison with StarGANv2-VC, we applied the JDC~\cite{JDC} model for $F_0$ and Parallel WaveGAN as vocoder~\cite{ParallelWaveGAN}. In our many-to-many evaluation, speakers from the training dataset were used, but with their new, unseen utterances. For the any-to-many scenarios, also known as unseen-to-seen settings, we used an utterance from the previously unseen speakers as input and translated it into the speaking style of the 20 reference speakers from training. To evaluate the performance of the models, we computed the FAD~\cite{FAD}, the Equal Error Rate (EER), and the perceptual Mean Opinion Score (pMOS) given by \cite{lo19_interspeech}. Further information about the metrics can be found in Appendix~(\wasyparagraph\ref{sec:metrics}). In all Tables, an average score is reported.
\subsection{Results}
\label{sec:results}
Qualitative results for many-to-many settings are shown in Appendix Table~\ref{tab:table-1}, and any-to-many in Table~\ref{tab:table-2}. The inference speed time is presented in Table~\ref{tab:table-3}. The FAD curve is shown in Appendix Figure~\ref{fig:FAD}. It can be seen in Table~\ref{tab:table-1} that our method provided highest results according to FAD metric, which is well-correlated with human judgment, making it effective in evaluating the quality of voice conversion systems~\cite{FAD}. However, our approach provides lower scores on the EER metrics and pMOS. But its important to note that our model has \underline{2x less} trainable parameters than DiffVC model, $\sim$\underline{25x faster} in inference, and trained only on 20 speakers. In comparison to StarGANv2-VC which is training via the weighted sum of $7$ different objectives, see Appendix (\wasyparagraph\ref{sec:baselines_appendix}), our method is training only via \underline{single} objective (\ref{eq:cond_dual-ot}) or (\ref{eq:extremal_cond_dual-ot}) aimed to find \textit{Wasserstein-2} transport map. In the next section we show how OT can be used to achieve high performance according to EER and pMOS as well. %This metric has its limitations, as it doesn't account for semantic relevance and it's sensitive to word sequence length differences. Moreover the EER results are dramatically based on the data pre-processing used during the training of the speaker verification model. % We suspect that this is because in our experiments no models were used to control the content results. %Also in comparison to the our approach StarGANv2-VC using an additional pre-trained joint detection and classification $F_0$ extraction network was employed to achieve $F_0$-consistent conversion.

\section{Experiments within Audio Latent Space Representation}
\label{sec:unit-experiments}
\subsection{Dataset and Baselines} For a fair comparison with kNN-VC, we used the LibriSpeech test-clean set and sampled 200 utterances, allowing for 5 per speaker. We converted each utterance to the remaining 39 speakers, resulting in a total of 7800 outputs per model. We compared our method to the kNN-VC as well as other any-to-any voice conversion systems including VQMIVC~\cite{wang2021vqmivc}, FreeVC~\cite{li2023freevc}, and YourTTS~\cite{casanova2022yourtts}. We applied the same settings as presented in kNN-VC experiments~\cite[Section 4]{baas2023voice}. 
\subsection{Settings} In addition to the metrics used in previous experiments, we also evaluated the word/character error rate (W/CER) of the converted speech using a pre-trained Whisper-base automatic speech recognition model~\cite{radford2023robust}, utilizing its default decoding parameters for transcription.\\
In line with the baseline kNN-VC, we utilized features extracted from layer 6 of WavLM-Large \cite{chen2022wavlm}, which generates a single vector for every 20 ms of 16 kHz audio. For SinkVC, we employed the Sinkhorn algorithm with a cost matrix determined by cosine similarity and entropy regularization set to 0.1. The four vectors with the highest scores in the recovered optimal plan were averaged to generate the resulting feature vector.  We did not engage in \textit{pre-matching training}~\cite{baas2023voice} of the vocoder in our experiments and used default Hifi-GAN provided by \cite{baas2023voice}. For FMVC, we separated the speaker's utterances into training and test categories. Subsequently, we trained the flow matching continuous map on pairs derived from the discrete Sinkhorn plan and tested on the unseen 100 utterance for the given speaker. A 3-layer MLP network was used to parameterize $v_\theta(t,x_t)$, with an additional input for time $t$ and a hidden size of 512. A batch size of 1000 feature vectors was used for 1000 iterations, alongside the Adam optimizer with a learning rate of $0.001$. %The learned flow matching was applied for each pair of speakers to test utterances.

\subsection{Results}
\label{sec:wav_results}
\begin{figure}[t!]
    \centering
    \includegraphics[width=7cm]{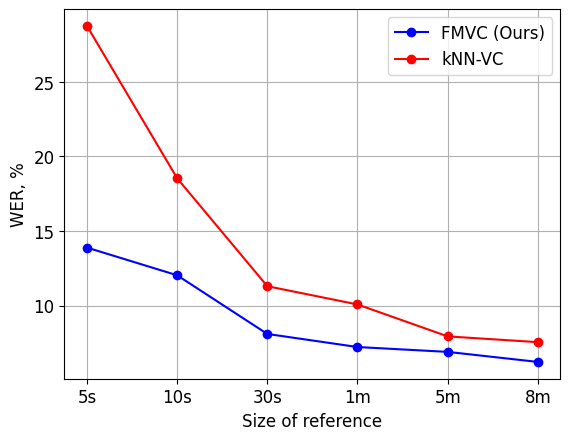}
    \caption{The figure displays the WER score in relation to the size of the target speaker's speech.}
    \label{fig:WER}
\end{figure}
\begin{table}[h!]
\centering
\small
\begin{center}
%\begin{small}
\begin{tabular}{l|lllll}
\hline
Model      & WER $\downarrow$    & CER $\downarrow$   & EER $\uparrow$ & FAD  $\downarrow$ & pMOS $\uparrow$ \\ \hline
VQMIVC  &59.46 & 37.55 &2.22 &  - &- \\
YourTTS  &11.93 & 5.51 &25.23 & - &-\\
FreeVC  &7.61 & 3.17 &8.87 & - & -\\
kNN-VC   & 7.54 & 3.41  & 40.5 & 2.92 & 3.72\\\hline
SinkVC   & 7.54 & 3.57  & \textbf{43.5} & 2.68 & 3.72 \\
FMVC  & \textbf{6.21} & \textbf{2.88}   & 32.5 & \textbf{2.50} & \textbf{3.77}  \\
\hline
\end{tabular}
%\end{small}
\end{center}
\caption{Any-to-any conversion results on the Librispeech dataset. The symbol $\uparrow$ indicates that a higher score is better, while $\downarrow$ indicates that a lower score is better.}
\label{tab:table-wav}
\end{table}
As can be seen in Table \ref{tab:table-wav}, our method consistently outperforms the kNN-VC predecessor across various metrics. To demonstrate that the continuous OT map provided by FMVC requires less data in the target domain, we conducted an ablation study on target size. For this, we applied VC and evaluated results using different metrics and various quantities of reference data (5s, 10s, 30s, 1m, 5m, 8m). As shown in Figure \ref{fig:WER}, our method yields results that are twice as effective as those of kNN-VC when using just 5 seconds of data, consistently outperforming on different target sizes. In Appendix (\wasyparagraph\ref{sec:wav_ablation}), we provide a visual representation for the other metrics, demonstrating superior performance as well. The EER scores are lower for FMVC, even though this method generates new points that may statistically differ from the data on which the ASR model was trained. Our method strikes a balance between complexity and performance, providing a \underline{continuous} solution that can map to \underline{new points} and consequently mitigate the limitations of its predecessor.
\section{Conclusion}
In contrast to StarGANv2-VC approach, our proposed method for mel-spectrogram representation is far simpler in terms of optimization, with its sole objective being to identify the \textit{Wasserstein-2} optimal transport map. We evaluated the quality of the speech conversion models using several automatic quality metrics (\wasyparagraph\ref{sec:results})(\wasyparagraph\ref{sec:wav_results}). Compared to DiffVC, our method demonstrated superior performance according to the FAD metric and proved to be more resource-efficient (as shown in Table~\ref{tab:table-3}). For the WavLM representation, we proposed a computationally lightweight FMVC approach (see Table~\ref{tab:table-wav}), which avoids the limitations of the SOTA any-to-any kNN-VC approach. We justify our algorithm by analyzing the recovered maps and showing that the learned map upper-bounded the FAD~\cite{FAD} between real and generated target data. We anticipate that our method will pave the way for future developments in the application of optimal transport in voice conversion tasks. 
%Our method surpasses previously proposed voice conversion techniques and exhibits strong results while being resource-efficient. 
% \subsection{Societal Impact}
% Conversion models can also be used by attackers to spoof voices and bypass security systems based on a particular person's voice characteristics (ASV for Automatic Speaker Veriﬁcation). On the other hand, conversion models can be used to improve the quality of ASV and eliminate vulnerabilities in such systems.
% % \subsection{Limitations}
% % As the limitation of the proposed approach we can consider the low scores achieved according to the EER metrics. Th used $\ell^2$ loss function do not include explicitly the content correspondence. to achieve an higher performance according to these metrics and still have a clear learning procedure, we can consider the proposed method in the latent space of the autoencoder with the accurately designed bottle neck, as mentioned in the AutoVC model.   

% \subsection{Reproducibility}
% To reproduce our experiment we provide source code in supplementary materials, run \texttt{experiments.sh} to start the training process. Details on used hyperparameters are presented in Settings (\wasyparagraph\ref{sec:settings}).
\bibliographystyle{IEEEtran}
\bibliography{mybib}
\newpage
\appendix
\onecolumn
\section{Analysis of the Recovered Map}
\label{sec:analysis}
%Tiven a pair ($\hat{f},\hat{T}$) that approximately solves equation (\ref{eq:cond_dual-ot}), it is natural to question the quality of the recovered optimal transport map $T$. In this subsection, we provide a bound on the difference between the optimal $T^*$ and $T$, which is based on the duality gaps for solving outer and inner optimization problems. For this, we only need to consider optimization over convex functions $f$, as seen in \cite[case 5.17]{villani2008optimal}. Our results below assume the convexity of $\hat{f}$, however, this may not hold in practice since $\hat{f}$ is typically a neural network.
Given a pair ($\hat{f},\hat{T}$) that approximately solves equation (\ref{eq:cond_dual-ot}), it is natural to question the quality of the recovered map $T$. To answer this, we provide a bound on the difference between the optimal $T^*$ and $T$, which is based on the duality gaps for solving outer and inner optimization problems. For this, we need to consider optimization over convex functions $f$, as shown in \cite[case 5.17]{villani2008optimal}. Our results assume the convexity of $\hat{f}$, however, this may not hold in practice, since $\hat{f}$ is typically a neural network.\\
\textbf{Theorem 1.} \textit{Assume that there exists a unique deterministic OT plan for quadratic cost between $\mu$ and $\nu$, i.e., $\pi^*$ for $T^{*}:\mathbb{R}^{D}\times\mathbb{R}^{S}\rightarrow\mathbb{R}^{D}$. Assume that $\hat{f}$ is $\beta$-strongly convex $(\beta>0)$ and $\hat{T}:\mathbb{R}^{D}\times\mathbb{R}^{S}\rightarrow\mathbb{R}^{D}$. Define:}
\begin{eqnarray*}
    &\epsilon_1 =\sup_{T} \mathcal{L}_{OT}(\hat{f},T) - \mathcal{L}_{OT}(\hat{f},\hat{T}) \text{ and } \\
    &\epsilon_2 =\sup_{T} \mathcal{L}_{OT}(\hat{f},T) - \inf_{f} \sup_{T} \mathcal{L}_{OT}(f,T).
\end{eqnarray*}
% \begin{equation}
%     \epsilon_1 =\sup_{T} \mathcal{L}_{OT}(f,G) - \mathcal{L}_{OT}(f,T) \text{ and } \epsilon_2 =\sup_{T} \mathcal{L}_{OT}(f,T) - \inf_{f} \sup_{T} \mathcal{L}_{OT}(f,T) 
% \end{equation} For the sampled condition $s\sim \eta$, corresponding samples from the source $x \sim \mu(\cdot|s)$ and target distributions $y \sim \nu(\cdot|s)$ are taken.
\textit{Then the following bound holds true for the $T^*$ from $\mu$ to $\nu$:}
\begin{eqnarray*}
    &\int_{\mathcal{S}}\frac{\text{FAD}(T_{\sharp}\mu(\cdot | s), \nu(\cdot| s))}{L^2} \leq \int_{\mathcal{S}} 2 \mathbb{W}^{2}_{2}(T_{\sharp}\mu(\cdot | s), \nu(\cdot| s))d\eta(s)\leq  \\
    &  \int_{\mathcal{X} \times \mathcal{S}} || T(x,s) - T^*(x,s)||d\mu(x,s) \leq \frac{2}{\beta}(\sqrt{\epsilon_1}+\sqrt{\epsilon_2})^2
\end{eqnarray*}\\
\textit{where FAD is the Frechet Audio Distance~\cite{FAD} and $L$ is the Lipschitz constant of the feature extractor of the pretrained neural network~\cite{FAD}}.\\
The duality gaps upper-bound the $\mathcal{L}^2(\mu)$ norm between the computed $T$ maps and the true $T^*$ maps, as well as the $\mathbb{W}_2$ between the true $\nu(\cdot|s)$ distributions and the generated $T_{\sharp}\mu(\cdot|s)$ distributions for all $s$.\\
\textbf{Proof}. Lets pick any $T^{*} \in \argsup_T \mathcal{L}(\hat{f}, T)=\argsup_T \int_{\mathcal{X}\times\mathcal{S}}\{\langle x, T(x,s)\rangle-\hat{f}(T(x,s),s)\} d \mu(x,s)$. 
% Or equivalently for any $x \in R^D$, $T^{*} \in \argsup_T \mathcal{L}(\hat{f}, T)=\argsup_y \int_{\mathcal{X}\times\mathcal{S}}\{\langle x, y\rangle-\hat{f}(y,s)\} d \mu(x,s)$. 

Consequently, for all $y \in R^D$ and $s \in R^D$:
\begin{equation}
    \langle x, T^{*}(x,s)\rangle-\hat{f}(T^{*}(x,s),s) \geq\langle x, y\rangle-\hat{f}(y,s)
\end{equation}
which, after regrouping the terms, results in
% \begin{equation}
%     \hat{f}(y,s) \geq \hat{f}(T^{*}(x,s))+\langle x, y\rangle- \langle x,T^{*}(x,s)\rangle
% \end{equation}
\begin{equation}
    \hat{f}(y,s) \geq \hat{f}(T^{*}(x,s),s)+\langle x, y-T^{*}(x,s)\rangle
\end{equation}
since $\hat{f}$ is $\beta$ strongly convex, for points $T(x,s)$, $T'(x,s)\in \mathbb{R}^D$ we derive.
\begin{equation}
    \hat{f}(T(x,s),s) \geq \hat{f}\left(T^{*}(x,s),s\right)+\left\langle x, T(x,s)-T^{*}(x,s)\right\rangle+\frac{\beta}{2}\left|T^{*}(x,s)-T(x,s)\right|^2.
\end{equation}
By regrouping the terms we get

\begin{equation}
    \left[\left\langle x, T^{*}(x,s)\right\rangle-\hat{f}\left(T^{*}(x,s),s\right)\right]-[\langle x, T(x,s)\rangle-\hat{f}(T(x,s),s)] \geq \frac{\beta}{2}\left|T^{*}(x,s)-T(x,s)\right|^2.
\end{equation}
Integration with respect to $x,s\sim \mu(x|s)\eta(s)$ gives us
\begin{equation}
    \epsilon_1=\mathcal{L}(\hat{f}, T^{*})-\mathcal{L}(\hat{f}, T) \geq \beta \int_{\mathcal{X}\times\mathcal{S}} \frac{1}{2}\left|T^{*}(x,s)-T(x,s)\right|^2 d \mu(x,s)=\frac{\beta}{2} \cdot\left|T-T^{*}\right|_{L^2(\mu)}^2.
\end{equation}
Lets $T^{*}$ be the OT map from $\mu$ to $\nu$. Using the fact that $T^{*}_{\sharp}\mu=\nu$.  
\begin{equation}
    \begin{array}{r}
\mathcal{L}\left(\hat{f}, T^{*}\right)=\int_{\mathcal{X}\times\mathcal{S}}\left\{\left\langle x, T^{*}(x,s)\right\rangle-\hat{f}\left(T^{*}(x,s),s\right)\right\} d \mu(x,s)+\int_{\mathcal{Y}\times\mathcal{S}} \hat{f}(y,s) d \nu(y,s)= \\
\int_{\mathcal{X}\times\mathcal{S}}\left\{\left\langle x, T^{*}(x,s)\right\rangle-\hat{f}\left(T^{*}(x,s),s\right)\right\} d \mu(x,s)+\int_{\mathcal{X}\times\mathcal{S}} \hat{f}\left(T^{*}(x,s),s\right) d \mu(x,s)= \\
\int_{\mathcal{X}\times\mathcal{S}}\{\underbrace{\left\langle x, T^{*}(x,s)\right\rangle-\hat{f}\left(T^{*}(x,s),s\right)+\hat{f}\left(T^{*}(x,s),s\right)}_{\geq\left\langle x, T^{*}(x,s)\right\rangle+\beta \frac{1}{2}\left|T^{*}-T^{*}\right|^2}\} d \mu(x,s) \geq \\
\int_{\mathcal{X}\times\mathcal{S}}\left\langle x, T^{*}(x,s)\right\rangle d \mu(x,s)+\beta \int_{\mathcal{X}\times\mathcal{S}} \frac{1}{2}\left|T^{*}-T^{*}\right|^2 d \mu(x,s).
\end{array}
\end{equation}
Let $f^{*}$ be an optimal potential in $\mathcal{L_{OT}}$ (6). Thanks to Lemma 4.2 according to \textit{Rout et. al, 2022} we can obtain that 
\begin{equation}
    \begin{array}{r}
\inf _f \sup _T \mathcal{L}(f, T)=\mathcal{L}\left(f^*, T^{*}\right)= \\
\int_{\mathcal{X}\times\mathcal{S}}\left\{\left\langle x, T^{*}(x,s)\right\rangle-f^*\left(T^{*}(x,s),s\right)\right\} d \mu(x,s)+\int_{\mathcal{Y}\times\mathcal{S}} f^*(y,s) d \nu(y,s)= \\
% \end{array}
% \end{equation}

% \begin{equation}
%     \begin{array}{r}
\int_{\mathcal{X}\times\mathcal{S}}\left\{\left\langle x, T^{*}(x,s)\right\rangle-f^*\left(T^{*}(x,s),s\right)\right\} d \mu(x,s)+\int_{\mathcal{X}\times\mathcal{S}} f^*\left(T^{*}(x,s),s\right) d \mu(x,s)= \\
\int_{\mathcal{X}\times\mathcal{S}}\left\langle x, T^{*}(x,s)\right\rangle d \mu(x,s).
\end{array}
\end{equation}
By combining the last two formulas (6) and (7) we get
\begin{equation}
    \epsilon_2=\mathcal{L}\left(\hat{f}, T^{*}\right)-\mathcal{L}\left(f^*, T^{*}\right) \geq \beta \int_{\mathcal{X}\times\mathcal{S}} \frac{1}{2}\left|T^{*}-T^{*}\right|^2 d \mu(x,s)=\frac{\beta}{2} \cdot\left|T^{*}-T^{*}\right|_{L^2(\mu)}^2.
\end{equation}
The inequality $\int_{\mathcal{X} \times \mathcal{S}} || T(x,s) - T^*(x,s)||d\mu(x,s) \leq \frac{2}{\beta}(\sqrt{\epsilon_1}+\sqrt{\epsilon_2})^2$ follows from the triangle inequality combined with (7) and (10). The inequality $\int_{\mathcal{S}} 2 \mathbb{W}^{2}_{2}(T_{\sharp}\mu(\cdot | s), \nu(\cdot| s))d\eta(s)\leq
\int_{\mathcal{X} \times \mathcal{S}} || T(x,s) - T^*(x,s)||d\mu(x,s)$ follows from the fact that for any $s$ the optimal map $T^{*}_{\sharp}\mu(\cdot|s)= \nu(\cdot|s)$ and \textit{Lemma A.2}[23]. 

Now, let $F$ be a VGGis model for extracting features from mel-spectograms, then the FAD score between the real and generated distributions is
\begin{equation}
    \int_{\mathcal{S}}\operatorname{FAD}\left(\hat{T}_{\sharp} \mu(\cdot|s), \nu(\cdot|s)\right)d\eta(s)=\int_{\mathcal{S}}\operatorname{FD}\left(F( \hat{T}_{\sharp} \mu(\cdot|s)), F(\nu(\cdot|s))\right)d\eta(s) \leq \int_{\mathcal{S}} 2\mathbb{W}_2^2\left(F(\hat{T}_{\sharp} \mu(\cdot|s)), F( \nu(\cdot|s))\right)d\eta(s)
\end{equation}
Where FD is the Frechet distance that lower bounds 2$\mathbb{W}_{2}^{2}$ following \textit{(Dowson and Landau 1982)}. Finally following \textit{Lemma 1} given in [23]:
\begin{equation}
    \int_{\mathcal{S}} \mathcal{W}_{2}^{2}\left(F(\hat{T}_{\sharp} \mu(\cdot|s)), F(\nu(\cdot|s))\right)d\eta(s) \leq \int_{\mathcal{S}} L^{2} \mathbb{W}_{2}^{2}\left(\hat{T}_{\sharp} \mu(\cdot|s), \nu(\cdot|s)\right)d\eta(s).
\end{equation}
Here L is the Lipschitz constant of $F$. Finally combining (9) and (10) we obtain
\begin{eqnarray*}
    &\int_{\mathcal{S}}\frac{\text{FAD}(T_{\sharp}\mu(\cdot | s), \nu(\cdot| s))}{L^2} leq int_{\mathcal{S}} 2 \mathbb{W}^{2}_{2}(T_{\sharp}\mu(\cdot | s), \nu(\cdot| s))d\eta(s)\leq 
    \int_{\mathcal{X} \times \mathcal{S}} || T(x,s) - T^*(x,s)||d\mu(x,s) \leq \frac{2}{\beta}(\sqrt{\epsilon_1}+\sqrt{\epsilon_2})^2
\end{eqnarray*} $\square$

\section{Mel-Spectogram Optimal Transport: Method and Evaluation Details}
\label{sec:details}
\subsection{Algorithm}
\label{sec:algorithm_appendix}
\begin{algorithm}[h!]
\SetInd{0.5em}{0.3em}
    {\SetAlgorithmName{Algorithm}{empty}{Empty}
        \SetKwInOut{Input}{Input}
        \SetKwInOut{Output}{Output}
        \Input{conditional distributions $\mu(\cdot|s),\nu(\cdot|s)$ for all $s$ accessible by samples, map $T_{\theta}$, potential $f_{\psi}$, cost $c(x,y)$; number of inner iterations $K_{T}$.}
        \Output{approximate OT map  $(T_{\theta})_{\#}\mu(\cdot|s)=\nu(\cdot|s)$}}
        \Repeat{not converged}{
            sample $s\sim\eta$; $x\sim\mu(\cdot|s)$, $y\sim \nu(\cdot|s)$;
            $\mathcal{L}_{f}\leftarrow  \frac{1}{|Y|}\sum\limits_{y\in Y}f_{\psi}(y,s) - \frac{1}{|X|}\sum\limits_{x\in X}f_{\psi}\big(T_{\theta}(x,s),s\big)$;
            update $\psi$ by using $\frac{\partial \mathcal{L}_{f}}{\partial \psi}$ to maximize $\mathcal{L}_{f}$;
            
            \For{$k_{T} = 1,2, \dots, K_{T}$}{
                sample $s\sim\eta$; $x\sim\mu(\cdot|s)$, $y\sim \nu(\cdot|s)$; ${\mathcal{L}_{T}\leftarrow\frac{1}{|X|}\sum\limits_{x\in X}\big[c
                \big(x, T_{\theta}(x,s)\big)- f_{\psi}\big(T_{\theta}(x,s),s\big)\big]}$;
            update $\theta$ by using $\frac{\partial \mathcal{L}_{T}}{\partial \theta}$ to minimize $\mathcal{L}_{T}$;
            }
        }
        \caption{Voice Conversion with Neural OT}
        \label{algorithm:main}
\end{algorithm}
\begin{table}[h!]
\centering
\begin{center}
%\begin{small}
\begin{tabular}{l|lllll}
\hline
      & StarGANv2-VC     & DiffVC    & NOT-VC(Ours)  \\ \hline
%AutoVC      & 10.9$\pm$0.05  & 0.18$\pm$0.02   & 3.51$\pm$0.2  \\
Time     &0.107 sec   &3.171 sec  &0.127 sec   \\
\hline
\end{tabular}
%\end{small}
\end{center}
\caption{Inference time for 5 seconds generation on \texttt{GPU Tesla V100-SXM3-32GB}.}
\label{tab:table-3}
\end{table}
\begin{figure}[h!]
    \centering
    \includegraphics[width=6cm]{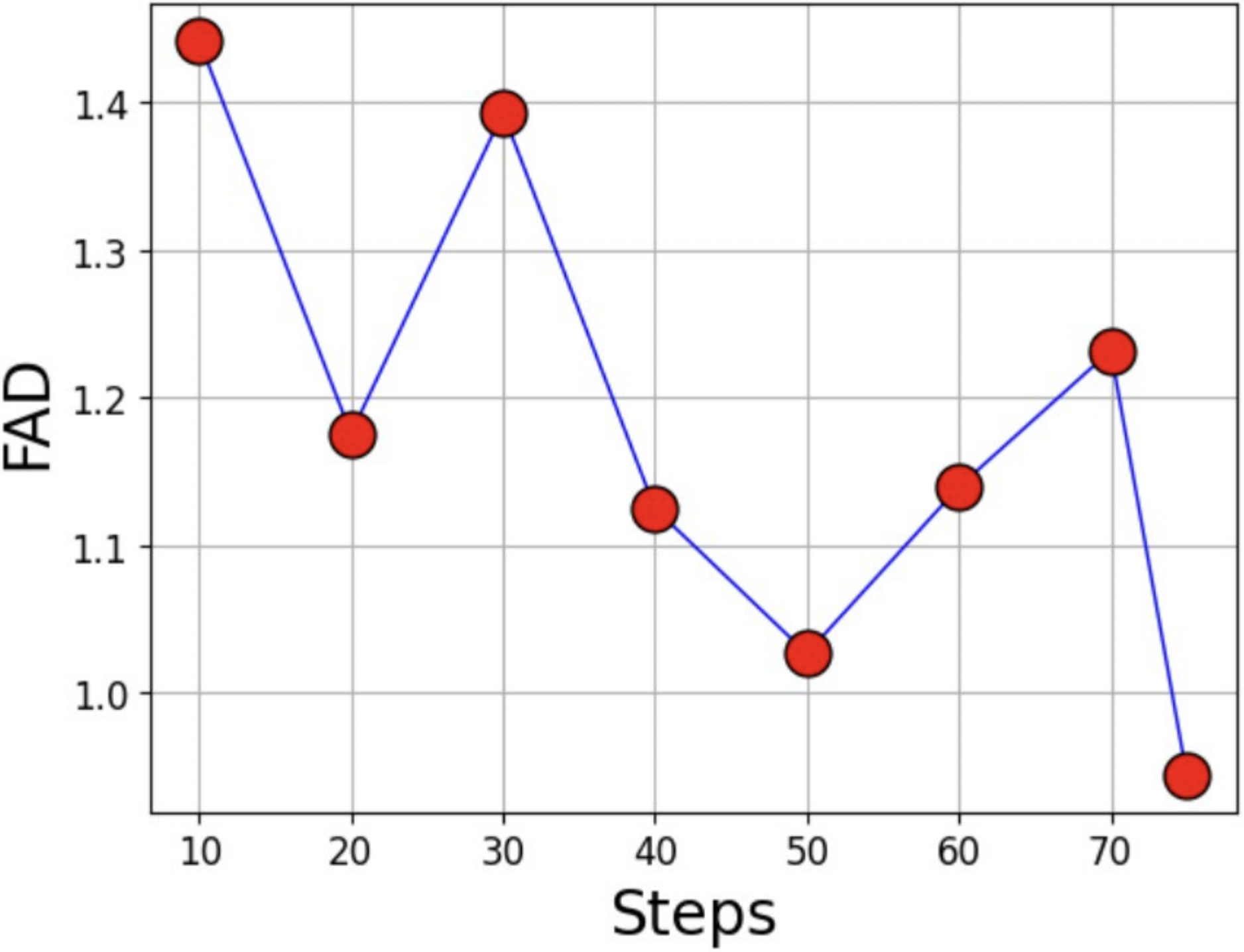}
    \caption{FAD scores for the many-to-many conversion problem, during training of our proposed XNOT-VC method. Steps are in 10e4 scale.}
    \label{fig:FAD}
\end{figure}
\subsection{Dataset Preprocessing}
\label{sec:details_data}
In our experiments we trained models on the common multi-speaker VCTK dataset ~\cite{veaux2016superseded}. This dataset contains recorded speech data from 109 native English speakers, amounting to approximately 44 hours of speech. Similar to StarGANv2-VC, we used only 20 speakers from the dataset. The implemented chalk spectrogram used a window of 1200 frames, a stepwise shift of 300 frames, and 80 chalk axis segments. The data was pre-processed similarly to StarGANv2-VC. All audio was resampled to 24,000 Hz, speech pauses longer than 100 ms were removed, audio per speaker was merged, and chunks of approximately five seconds were created. During training, chunks with an approximate length of 2.5 seconds were randomly selected. In all experiments the data is non-parallel, i.e the different contend is in the source and target speaker. We tested the methods in many-to-many and any-to-many settings(\wasyparagraph\ref{sec:settings}). For the fair comparison with the StarGANv2-VC, we utilized the JDC~\cite{JDC} model for $F_0$ and Parallel WaveGAN as vocoder~\cite{ParallelWaveGAN}.
\subsection{Related Work and Baselines}
\label{sec:baselines_appendix}
To compare our method we considered a different methods that solves many-to many voice conversion training, in which, each speaker is treated as an individual domain. Importantly the data is non-parallel, i.e the different contend is in the source and target speaker. We tested the methods in many-to-many and any to many settings.\\
\textbf{AutoVC.} First of all we compare our method to the currently the simplest approach for the VC~\cite{Qian2019}. Auto VC provides a style transfer scheme that involves only an autoencoder loss with a carefully designed bottleneck size. The WaveNet~\cite{oord2016wavenet} vocoder is used to transform spectograms back into the raw audio. \\
\textbf{DiffVC}: We also considered a comparison with the diffusion-based VC model~\cite{popov2021diffusion}. The generator network of this model consists of 127M parameters. This model receives a mel-spectrogram and,by using 30 diffusion steps, converts the voice style of the source speaker. We compared our method with the pre-trained \textit{Diff-VCTK} model provided by the authors. The Hifi-GAN~\cite{kong2020hifi} was used for the final audio generation. \\
\textbf{StarGANv2-VC}: Finally we compared our method to the StarGANv2-VC\cite{Li2021}. This architecture based on a discriminator and generator. The generator converts a mel-spectrogram into the target speaker spectrogram, using the style encoder output $h$ and the $F_0$ network outputs $G(\mathcal{X}, s, F_0)$. To avoid confusion, it is important to note that there exists a similar model named \textit{StarGAN-VC2}~\cite{Li2021}. Although the idea behind both models are the same, StarGANv2-VC shown better results. StarGANv2-VC operate directly in the input space, which allows the preservation of sample's intrinsic structure and ensures generator network validation. However, these methods requires complex training strategies. Specifically, these methods involve a range of training objectives with an loss-specific weight coefficient $\lambda$. For example, the total objective $\mathcal{L}_{GAN}$ for the StarGANv2-VC~\cite{Li2021} model can be written as:
\begin{eqnarray}
     &  \mathcal{L}_{GAN} = \lambda_{adv}\mathcal{L}_{adv}+\lambda_{sty}\mathcal{L}_{sty} + \lambda_{cyc}\mathcal{L}_{cyc} + \lambda_{norm}\mathcal{L}_{norm} - \lambda_{div}\mathcal{L}_{div} + \lambda_{F0}\mathcal{L}_{F0} + \lambda_{asr}\mathcal{L}_{asr}.
\end{eqnarray}
Where $\mathcal{L}_{adv}$ represents adversarial loss that renders the converted feature indistinguishable from the real target feature via GAN discriminator~\cite{goodfellow2014generative}. The style restoration loss or classification loss ($\mathcal{L}_{sty}$) aims to obtain a style embedding approximately equal to the target audio. Identity loss ($\mathcal{L}_{id}$) preserves properties of the input audio. The relative pitch loss ($\mathcal{L}_{F0}$) controls the ratio of the frequency at which a person speaks. Linguistic content loss ($\mathcal{L}_{asr}$) preserves the speech content by minimizing the distance in the hidden layer of the pretrained Automated Speech Recognition (ASR)~\cite{malik2021automatic} model. Finally, cyclical loss ($\mathcal{L}_{cyc}$) is also used to penalize the generator if converted audio turned again into the reference one is not equal to the input one. The incorporation of all these losses into training \textit{complicates the optimization and requires an intensive hyper-parameter search} on $\lambda$. Hence, the convergence property of GAN-based VC models is fragile~\cite{Qian2019}.

We compared our solution to the pre-trained models on the VCTK dataset provided by the authors~\cite{Li2021}. The Parallel WaveGAN~\cite{yamamoto2020parallel} was used for the audio generation.
\subsection{Metrics}
\label{sec:metrics}
\textbf{FAD}~\cite{FAD}:Calculates the distance between the distribution of the real and converted voices. FAD is based on the audio embedding, from a pre-trained audio or speech recognition model on the real and converted speech signals. Lower FAD scores indicate that the converted voice is closer to the distribution of the real voice. We calculated the metric value separately for each speaker, and then calculated the average value for all speakers. As the original data, we calculated statistics on the training dataset, for the generated data, we randomly sampled 1000 audio examples from the test data and converted to the target speaker.\\
% It quantifies both the quality of generated audio as well as the level of preservation/removal of speaker identity. 
\textbf{EER}: Equal Error Rate is calculating the False Acceptance Rate and False Rejection Rate of the voice recognition system  and find the point at which they are equal. We chose WavLM Base+~\cite{Chen2021WavLM} as the speaker verification model for all experiments. The Equal Error Rate provides a single value that balances these two rates against each other. In voice conversion models, an EER close to 0 would represent a high-performing model.\\ %For each selected example in test set, we referenced an example from the corresponding speaker in the training set. We selected another example from the test set, this time from a different speaker, for conversion to the tested speaker. This conversion, combined with sampling of an example from the training set, produced negative examples.
\textbf{pMOS}: The Perceptual Mean Opinion Score is represents the average ratings given by human evaluators. Evaluators rate the quality of the voice on a scale, typically from 1 (worst) to 5 (best). Using the provided human demonstrations, neural approximation is built. Higher pMOS indicating a better performance. In experiments we used model proposed in~\cite{lo19_interspeech}. %
\subsection{Training} 
\label{sec:details_models}
For the NOT-VC and XNOT-VC, transport map network $T_\theta$ with 46.0 M parameters UNET neural network was used. For the potential $f_\omega$ 55.1 M parameters model was utilized.  The optimization parameters: AdamW~\cite{AdamW} was chosen as the optimizer for both the map $T$ and the potential $f$, the learning rate equal to $5\cdot 10^{-5}$, weight decay was chosen to be $1\cdot 10^{-10}$. Learning rates of generator and potential is equal to 0.001. The number of generator optimization steps $K_T$ in algorithm\ref{algorithm:main} was set to 10 per one potential $f$ parameters update. The batch size was equal to 5. The code is written using \texttt{PyTorch} framework and will be made publicly available. We trained two models, model with the loss (\ref{eq:cond_dual-ot}) called NOT-VC, and the model trained in the eXtremal formulation (\ref{eq:extremal_cond_dual-ot}) with the parameter $w=12$ called XNOT-VC.\\ %The \textit{cost} function in these settings is used exclusively to store the \textit{content}, while the potential $f$ catches only the \textit{style}. 
Additionally, to show that our approach is suitable for learning a mapping with a loss other than $\ell^2$, we have provided additional experiments using the ASR-based cost function. %All settings for the ASR based transport were the same as those provided in (\wasyparagraph\ref{sec:settings}). In this experiment, we explicitly separated content and style learning into two parts. 
% \begin{figure}[t!]
%     \centering
%     \includegraphics[width=2cm]{AISTATS2023PaperPack/Images/TSNE.png}
%     \caption{TSNE}
%     \label{fig:TSNE}
% \end{figure}

\subsection{Testing} 
\label{sec:settings}
We used two evaluation pipelines: the many-to-many and the any-to-many settings. In the many-to-many settings, we evaluated the performance of the models using the input and reference speakers from our training dataset, but on the new utterances of these speakers. In the any-to-many settings (often referred to as unseen-to-seen), we used as input an utterance from the a previously unseen speakers and converted it to the speaking style of the 20 reference speakers used in training. We computed the FAD metric by sampling 1000 examples from the target speakers. We then randomly selected 100 input samples from speakers in the test set. Each new sample was then converted to match the speech style of 10 reference speakers. We calculated the FAD by comparing the real target data with the generated data and reported \textit{the average.}\\
In the any-to-many setting, we followed the same procedure, but selected 100 samples from speakers not presented in the training set. We used the same data for the pMOs metric. To compute the EER, we sampled 50 unseen audio clips from each speaker in the training set. For each of these clips, we converted it to match all reference speaker sets, for a total of 20. Then we randomly sampled one audio clip from the training set for the reference speaker, called the target audio. This process resulted in 50 * 20 scores for the generated audio and 50 * 20 scores for the original audio.

\section{WavLM Optimal Transport: Method and Evaluation Details}
\subsection{Reference size ablation}
\label{sec:wav_ablation}
% \begin{figure}[h!]
%     \centering
%     \includegraphics[width=6cm]{Images/pmos_vs_time.png}
%     \caption{pMOS scores in depending of the provided target len.}
%     \label{fig:pMOS}
% \end{figure}
\begin{figure}[h!]
    \centering
    \includegraphics[width=7cm]{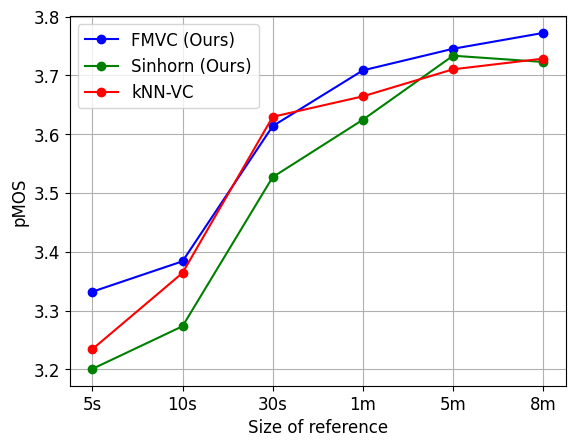}
    \caption{pMOS scores in depending of the provided target len.}
\end{figure}
\begin{figure}[h!]
    \centering
    \includegraphics[width=7cm]{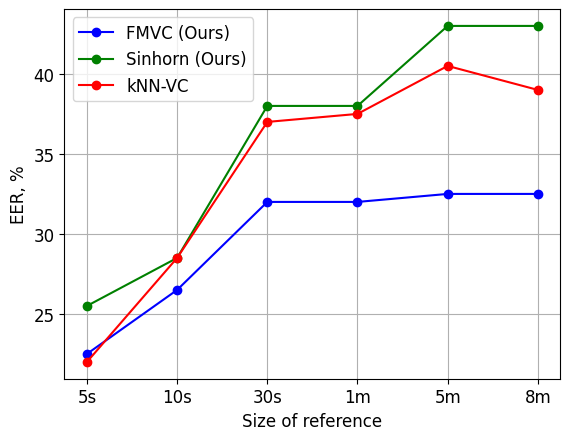}
    \caption{EER scores in depending of the provided target len. EER scores is lower for FMVC, while this method generates a new score that may be statistically different from the data on which EER ASR model was trained.}
\end{figure}
\begin{figure}[h!]
    \centering
    \includegraphics[width=7cm]{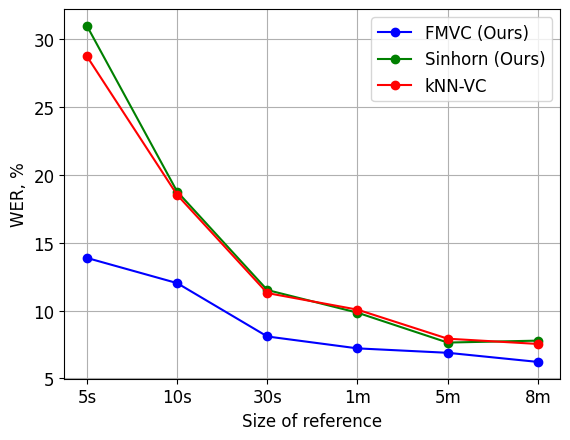}
    \caption{WER scores in depending of the provided target len.}
\end{figure}
\end{document}